\documentclass[12pt]{article}
\usepackage{graphicx}
\usepackage{subfigure}
\usepackage[dvips]{color}
\usepackage{amssymb}
\usepackage{amsmath}

\input epsf.tex

\newcommand{\beq}{\begin{equation}}
\newcommand{\eeq}{\end{equation}}
\newcommand{\be}{\begin{equation}}
\newcommand{\ee}{\end{equation}}
\newcommand{\bea}{\begin{eqnarray}}
\newcommand{\eea}{\end{eqnarray}}

\makeatletter

\@addtoreset{equation}{section}
\makeatother

\def\href#1#2{#2}

\begin{document}

\baselineskip=15.5pt
\pagestyle{plain}
\setcounter{page}{1}

\begin{titlepage}
\begin{flushleft}
       \hfill                       FIT HE - 12-03 \\
       \hfill                       KYUSHU-HET 138 \\
\end{flushleft}

\begin{center}
  {\huge Holographic cold nuclear matter   \\ 
   \vspace*{2mm}
and neutron star  \vspace*{2mm}
}
\end{center}

\begin{center}

\vspace*{2mm}
{\large Kazuo Ghoroku${}^{\dagger}$\footnote[1]{\tt gouroku@dontaku.fit.ac.jp},
Kouki Kubo${}^{\ddagger}$\footnote[2]{\tt kkubo@higgs.phys.kyushu-u.ac.jp},
Motoi Tachibana${}^{Q}$\footnote[3]{\tt motoi@cc.saga-u.ac.jp},
\\
and Fumihiko Toyoda${}^{\P}$\footnote[4]{\tt ftoyoda@fuk.kindai.ac.jp}
}\\

\vspace*{2mm}
{${}^{\dagger}$Fukuoka Institute of Technology, Wajiro, 
Higashi-ku} \\
{
Fukuoka 811-0295, Japan\\}
{
${}^{\ddagger}$Department of Physics, Kyushu University, Hakozaki,
Higashi-ku}\\
{
Fukuoka 812-8581, Japan\\}
{${}^{Q}$Department of Physics, Saga University, Saga 840-8502, Japan\\}
{
${}^{\P}$Faculty of Humanity-Oriented Science and
Engineering, Kinki University,\\ Iizuka 820-8555, Japan}

\vspace*{3mm}
\end{center}

\begin{center}
{\large Abstract}
\end{center}
We have previously found a new phase of cold nuclear matter based on a holographic gauge theory, 
where baryons are introduced as instanton gas in the probe D8/$\overline{\rm D8}$ branes. 
In our model, we could obtain the equation of state (EOS) of our nuclear matter by introducing
fermi momentum. Then,
here we apply this model to the neutron star and study its mass and radius by solving 
the Tolman-Oppenheimer-Volkoff (TOV) equations in terms of the EOS given here.
We give some comments for our holographic model from a viewpoint of the other 
field theoretical approaches.

\noindent

\vfill
\begin{flushleft}

\end{flushleft}
\end{titlepage}
\newpage

\vspace{1cm}

\section{Introduction}

In a holographic approach, 
the baryon is introduced as a soliton of vector mesons 
constructed in the probe $N_f$ flavor branes
\cite{SS,HSSY,Hashimoto:2008zw,Hashimoto:2009ys}. 
The configuration of this
soliton, which has a unit baryon number, is given by a finite-sized BPST
instanton solution of the $SU(N_f)$ ($N_f\geq 2$) YM theory \cite{BPST} in D8 branes. 
Here the size of the instanton
is not arbitrary and
determined to be as a finite size since the solution is embedded
in the non-flat 4D partial space of the probe brane \cite{HSSY}. 
In this context, several approaches to study the nuclear system
have been performed 
\cite{Kim:2006gp,BLL,Kim:2007vd,RSRW,Chuang:2010ku,Kaplunovsky:2012gb,Seki:2012tt,deBoer:2012ij}.

Previously we have studied nuclear matter according to this holographic approach \cite{GKTTT}. 
In our approach, the flavored Yang-Mills fields in the action of D8 branes are retained  
up to the quadratic term of the field strength. 
Its higher order terms are neglected here in the context of our dilute gas approximation of the instantons.
The configuration of the flavored YM fields is obtained by supposing the dilute instanton
gas form and by determining the size 
parameter of the instantons from the action principle \cite{GKTTT}. After that, 
the chemical potential $\mu$ and the charge density $\bar{n}$ for the baryon
are obtained in terms of the solution of the $U(1)$ gauge field, which
is dual to the baryon number current operator. In this process, another important
point is to find the embedding solution for the profile of D8/$\overline{\rm D8}$ brane. 
As for this profile, we use an
antipodal solution \cite{HSSY}, which is originally obtained in the case
without any instantons. However, we could show that our antipodal
solution, which is shown in \cite{GKTTT}, 
is also useful for any instanton (baryon) density $n$\footnote{The charge density 
$\bar{n}$ is defined through the time component of $U(1)$ gauge field. While it
is proportional to the baryon number $n$, we use $n$ in our present analysis.}
\cite{GKTTT}.

In this setting, we could find an interesting phenomenon, a phase transition of the nuclear matter,
which has been shown by
$\bar{n}$-$\mu$ relation. Namely, $\bar{n}$
jumps from zero to a finite
value $\bar{n}_c$ at a non-zero $\mu=\mu_c$. This corresponds to a transition from the
vacuum to a non-trivial nuclear matter phase. 
It would be important to study the properties of this new phase of the nuclear matter and to
find a possible candidate for this matter.

\vspace{.5cm}
Our purpose 
is to exploit our analysis for $\bar{n}$-$\mu$ relation
in order to identify the nuclear matter given here with a star. 
Since the baryon considered here is neutral, we could suppose such a star as a neutron star. 
It is an interesting problem to see that this proposal is reasonable or not
by solving the TOV equations \cite{TOV}. 
In solving the TOV equations however, we must know the relation of
the energy density $\rho$ and pressure $p$ of the nuclear matter, namely EOS. 
In general, $\rho$ and $p$ are holographically obtained by using the bulk metric 
according to holographic renormalization \cite{KS,KSS}. 
In the present case, however, the bulk metric is 
independent of the instanton gas since the flavor branes are treated as the probe.

{An approach to solve the TOV equations
has been performed by using a holographic EOS obtained 
in a context of holographic framework \cite{KLSW}, however it leads to a 
large radius star.}
Then, we propose an alternative way to get the relation of $\rho$ and $p$, EOS.
We know the Luttinger theorem for interacting fermion system \cite{La}. We apply this 
theorem to our nuclear matter. Then we acquire the other formula satisfied by the
usual fermion system. One exception is the
point that the mass of the fermion in our model is introduced
as a function of the baryon number density. This density dependence of the fermion mass
implies a non-trivial interaction among the fermion system given here, and this
density dependent mass is determined from our holographic model. After that, we
obtain a EOS of our holographic model, then the TOV equations\cite{TOV} are solved.

\vspace{.3cm}
The outline of this paper is as follows. In the next section, the instanton gas model is set up and how to obtain the EOS is explained. In the section 3, the holographic EOS is given from
our analysis, then it is used in the next section to solve the TOV equations. We give some
speculation on our result, which give a rather small mass and radius for the neutron star,
and an estimation based on the two phase components is given.
In the final section, summary and discussions are given.

\section{Instanton gas and EOS}\label{sec2}

\subsection{Instanton gas }\label{subsec2-1}

Here baryon is introduced as an instanton soliton found in the $N_f$ 
D8 branes, which are embedded
as probe in the D4 stacked bulk \cite{HSSY} in the type IIA string theory.
We set as $N_f=2$, for simplicity, then the $U(2)=U(1)\times SU(2)$ gauge fields in the D8 brane
action are given as follows
\bea
   A_b&=&A_b(z)\delta_b^0\, ,\\
   (\vec{f}_1)_{ij}&=&Q(x^m-a^m,\rho_I)\epsilon_{ijk}\tau^k\, ,\label{an1} \\
   (\vec{f}_1)_{iz}&=&Q(x^m-a^m,\rho_I)\tau^i\, ,\label{an2}
\eea
where $A_b$ ($(\vec{f}_1)_{ij}$, $(\vec{f}_1)_{iz}$) denotes the $U(1)$ gauge field
($SU(2)$ gauge field strength). Further,
$\rho_I$ ($a^m$) denotes the instanton size (position), $\epsilon_{123z}=1$, 
$i,j=1,2,3$ and  $m=1,\dots ,4$, where $x^4=z$.
Then, $Q$ is extended into the following multi-instanton form, 
\beq\label{dilute-gas-approximation}
 Q=\sum_{i=1}^{N_I} Q_i\, , \quad Q_i={2\rho_{I_i}^2\over \left((x^m-a^m_i)^2+\rho_{I_i}^2\right)^2}\, ,
\eeq
where $N_I$ denotes the instanton number.
By neglecting the higher order terms of $Q$ and the overlapping of instantons,
the $U(1)$ gauge field and the instanton size are determined as a solution of the brane
action with the Chern-Simons term. The chemical potential of the baryon is obtained
by using the result for $E_z=\partial_zA_0$ as \cite{GKTTT} 
\beq
   \mu =\mu_c+\int_0^{\infty}dz E_z\, , \label{chemical-pot} 
\eeq
where $\mu$ and $\mu_c$ are defined as
\beq
  \mu=A_0(\infty)\, , \quad \mu_c=A_0(0)\, .
\eeq

After that, we can obtain the relation between the
instanton density, $n=N_I/V_3$ where $V_3$ is the three dimensional volume,
and the chemical potential $\mu$ of the baryon. From this
result, we can derive the equation of the nuclear matter given here. 
The details of our model are given in \cite{GKTTT}.

\subsection{Density dependent mass and EOS}

As mentioned in the introduction, we apply the Luttinger theorem for interacting 
fermion system \cite{La} to our nuclear matter.
Then we may write the baryon number density $n$ as 
\beq
n=\frac{gS_{2}}{3(2\pi)^3}k_F^3={8\pi\over 3(2\pi)^3}k_F^3 \label{n_kf}
\eeq 
by introducing the fermi momentum $k_F$ for our baryon. $g(=2)$ is a
number of spin degrees
of freedom of the baryon, $S_2$ is a volume of two-dimensional surface of 3-sphere.
In order to exploit this relation and our result $\mu=\mu(n)$, we introduce an effective
baryon mass $m(n)$, which depends on the density $n$. This effective mass is interpreted as
a reflection of the complicated interaction among the nuclear system given here. A similar concept
is seen in the ordinary field theory approach \cite{Schmitt}.
Then it is possible to express $\mu$ as
\beq
  \mu=\mu(n)=\sqrt{k_F^2+m^2(n)}\, . \label{mu_kf}
\eeq
In this context, by extending the formula for the free fermion gas, 
we get the energy density $\rho$ and pressure $p$ of the nuclear matter
are obtained as,
\bea
  \rho &=& \rho_c+{8\pi\over ( 2\pi )^3}\int_{k_c}^{k_F} dk~k^2\sqrt{k^2+m^2(n)}\, , \label{energy} \\
  p &=& p_c+{8\pi\over 3(2\pi )^3}\int_{k_c}^{k_F} dk~{k^4 \over \sqrt{k^2+m^2(n)}}\, , \label{press}
\eea
where $\rho_c$ and $p_c$ denote the critical density and pressure respectively.
Here notice that the lower bound of the momentum, $k_0$, is introduced as
\beq
   n_c={8\pi\over 3(2\pi)^3}k_c^3
\eeq
since the nuclear matter considered here is defined for $n\geq n_c$.
This is our proposal to obtain the energy momentum tenser of the nuclear matter based on the
holographic approach. Then we obtain EOS, the relation between $\rho$ and $p$, at any density to solve the TOV.

\section{Numerical Data of $\mu(n)$ and $m(n)$ and EOS}
Now, we reconstruct the information of density dependence of an effective nucleon mass $m\left(n\right)$ in nuclear matter phase from our $\left(\mu,n\right)$
data\footnote{Data points are limited in the range of $\langle r\rangle >$ nucleon size where $\langle r\rangle$
is the separation among nucleons .The dilute gas approximation is available.}
 by using the eqns. (\ref{n_kf}) and (\ref{mu_kf}).
\par
\begin{figure}
\includegraphics{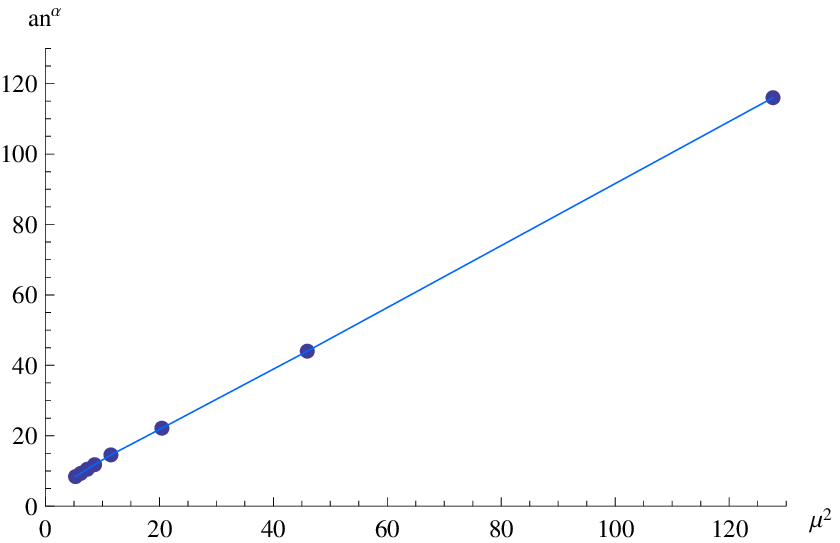}
\includegraphics{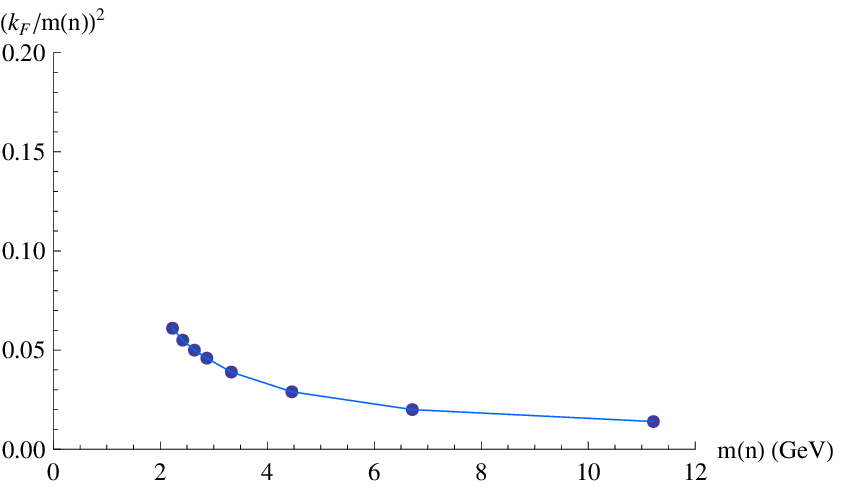}
\caption{Left:The relationship between $\mu^2$ and $an^{\alpha}$.
Right:The relationship between $(k_F/m(n))^2$ and $m(n)$.  $a=1500, \alpha=1.0
$.  $n/n_0=4\sim 50$ with $n_0=1.36\times 10^{-3} \rm{GeV^3}$(normal nuclear matter).
\label{fig:n-mu}}
\end{figure}

From Fig.2 of the previous paper \cite{GKTTT}, we get the empirical relation
\begin{eqnarray} 
\mu^2 \sim a n^{\alpha}
\end{eqnarray}
with $a=1500, \alpha=1.0$ in natural unit (mass unit is set to GeV).
This is shown in Fig.\ref{fig:n-mu}.left.
As $a>>1$, we can estimate $\mu^2>> k_F^2$ and $\mu(n)\sim m(n)$ from the eqns. (\ref{n_kf}) and (\ref{mu_kf}) which is shown in  Fig.\ref{fig:n-mu}.right.  We obtain $m(n)$ and see the relation 

\begin{eqnarray}
 m^2(n)\sim \mu^2(n) \sim a n^{\alpha} =a({8\pi\over3(2\pi)^3}k_F^3)^{\alpha}
\end{eqnarray}

\par
The value of the fermi momentum and the mass $m_0$ at
phase transition point are $k_c\equiv
\left(\frac{3\left(2\pi\right)^{3}}{8\pi}n_c\right)^{1/3}\sim0.55$,
$m_0^2\equiv m^2(n_c)\sim 4.99$.
\par

Next, we calculate the fermi momentum dependence of the energy
density and the pressure from eqns.(\ref{energy}) and (\ref{press}).
\par
As $k \leq k_F << m(n)$,the energy function $\sqrt{k^2+m^2(n)}$
can be approximated as

\begin{eqnarray}
  \sqrt{k^2+m^2(n)} \sim m(n)
\end{eqnarray}
 in the integral.
\par
Then we can calculate the integrals in eqns.(\ref{energy}) and (\ref{press}),
\begin{eqnarray}
\rho\left(k_{F}\right) & = & \rho_{c}+b_{1}\left(k_{F}^{3\alpha/2+3}-k_{c}^3k_F^{3\alpha/2}\right),\label{eq:rho_in_nuclear_matter}\\
p\left(k_{F}\right) & = & p_{c}+b_{2}\left(k_{F}^{5-3\alpha/2}-k_{c}^5k_F^{-3\alpha/2}\right),\label{eq:p_in_nuclear_matter}
\end{eqnarray}
where $\rho_{c}\equiv\rho\left(k_{c}\right),\ p_{c}\equiv p\left(k_{c}\right)$ and
$$b_{1}=\frac{\sqrt{a\left(\frac{1}{3\pi^{2}}\right)^{\alpha}}}{3\pi^{2}}, \quad b_{2}=\frac{1}{15\pi^{2}\sqrt{a\left(\frac{1}{3\pi^{2}}\right)^{\alpha}}}$$.

\begin{figure}
\includegraphics{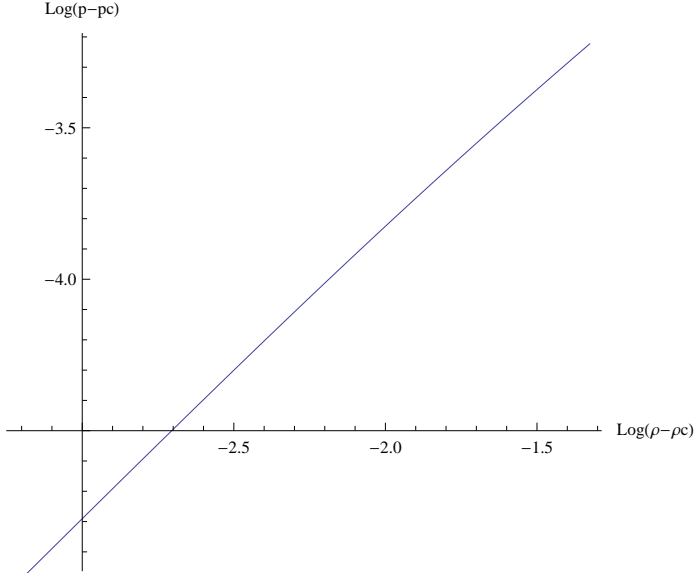}
\caption{The relationship between $\rho$ and $p$. It is approximated as
 $p-p_c=\mathrm{const.}\times(\rho-\rho_c)^\gamma$ with $\gamma=0.78$.
\label{fig:p-rho}}
\end{figure}

From eqns.(\ref{eq:rho_in_nuclear_matter}) and (\ref{eq:p_in_nuclear_matter}), we obtain the relation as 
\beq\label{eq:EoS}
\rho\left(p\right)=\rho_{c}+b_{1}\left[\left(\frac{1}{b_{2}}\left(p-p_{c}\right)+k_{c}^5k_F^{-3\alpha/2}\right)^{\frac{3\alpha/2+3}{5-3\alpha/2}}-k_{c}^3k_F^{3\alpha/2}\right]\
.
\eeq
Parameters $a(=1500)$ and $\alpha(=1.0)$ lead to $b_1=0.239$ and $b_2=0.952\times 10^{-3}$.
Fig.\ref{fig:p-rho} shows the relationship between $\rho$ and $p$.
\par
As $1/b_2 >>1$ ,this equation is approximated by the following EOS

.
\beq\label{eq:EoS3}
p-p_{c}={b_2\over b_1^{\gamma}}(\rho-\rho_{c})^{\gamma}
.
\eeq

The index $\gamma(={ 5-3\alpha/2 \over 3\alpha/2+3}=0.78<1)$ is abnormal. This is in contrast to normal stars($\gamma=4/3$) or the ideal fermion gas($\gamma=5/3$). The origin of $\alpha=1$ is the effective nucleon mass $m(n)$. As the nucleon density increases, the effective nucleon mass and nucleon size
 increase in our holographic model \cite{GKTTT}. 

\newpage
\section{Application to Neutron Star and solution of TOV Equations}
Since we have EOS as eq.(\ref{eq:EoS3}), we can reduce a unknown function in TOV equations \cite{TOV} and can solve them.
The TOV equations are
\par
\begin{eqnarray}
{dp(r) \over dr}&=&{G(m(r)+4\pi r^3p(r)/c^2)\over c^2r^2}{c^2\rho(r)+p(r)\over 1-2Gm(r)/c^2r}, \nonumber \\
 {dm(r) \over dr}&=&4\pi r^2\rho(r),
\end{eqnarray}
where $G$ is the Newton's constant, $p(r)$ and $m(r)$ are the pressure and the mass respectively.
We introduce new parameters $ r_0, M_0, p_0$ as follows,
\begin{eqnarray}
 \frac{c^2}{G} &=& \frac{M_0}{r_0}   \nonumber  \\
r_0&=&1  \rm{~km}, \nonumber \\
M_0 &=&1.35\times 10^{30} \rm{~kg}(=0.68\times M_s(\rm{solar mass})), \nonumber \\
p_0 &=&{M_0\over r_0^3}c^2(=1.22\times 10^{38} \rm{~N/m^2}). \nonumber
\end{eqnarray}
The TOV equations can expressed by the dimensionless values  $x=r/r_0,  Y(x)=m/M_0,  Z(x)=p/p_0$,
\begin{eqnarray}
 {dZ(x)\over dx}x^2(1-2Y(x)/x)=(Y(x)+4\pi x^2 Z(x))({dY(x)\over dx}/4\pi x^2+Z(x)) .
\end{eqnarray}
\par
When we combine the EOS  and the TOV equations, we can estimate the radius and mass of neutron star depending on parameters of $\rho_c$ and $p_c$.
\par
\vspace{0.5 cm}
(1){\bf  the case of $p_c=0$}.
\par
\begin{figure}[htbp]
\begin{center}
\includegraphics[width=9.0cm,height=6cm]{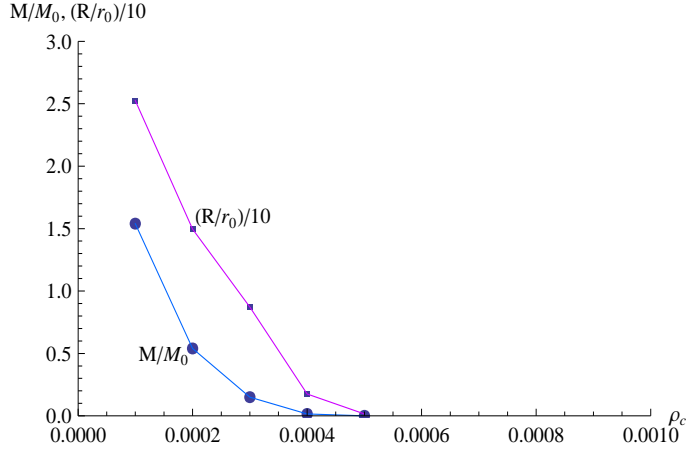}
\caption{{\small The radius and mass of neutron star vs  $\rho_c$ in the case of $p_c=0$. }}
\label{TOV-1}
\end{center}
\end{figure}

This case is interesting  and  neutron star is composed of the nuclear matter in new phase.
Calculated radius and mass  depend on the parameter $\rho_c$ and are shown in Fig.\ref{TOV-1}.
\par

If we can select $\rho_c\simeq 0.0002 \rm{~GeV^4}$, we obtain the neutron star of $R\simeq 20 \rm{~km}$ and $M\simeq M_s$.
\par
When  $\rho_c > 0.0005 \rm{~GeV^4}$, the radius and mass are quite small.
\par
As $\rho_c \rightarrow 0$, $R,M \rightarrow \infty$ and the neutron star becomes unstable\cite{KLSW}.

\par
{\bf the meaning of  $\rho_c$}
\par
 We have found that the resultant mass and radius of neutron star depend on the critical energy density $\rho_c$.
What is the natural value of $\rho_c$? A naive expectation is that $\rho_c \approx m_B n_{cr}$. In this case,
$\rho_c$ is approximately given as 0.005 GeV$^4$ and then both the mass
and radius get quite small.

The following consideration may be useful to understand our results. Suppose that a neutron star is composed
by $A$  neutrons, where mass and mutual distance are given as $m_B$ and $d_0$. Then the neutron star mass $M_{NS}$ and
radius $R_{NS}$ will be
\begin{eqnarray}
M_{NS} \approx m_B A, \qquad R_{NS} \approx d_0 A^{1/3}.
\label{NS mass-radius}
\end{eqnarray}
If we supposed that the neutron star is the star which has the smallest
radius not to be the black hole, the radius should agree with the
Schwarzschild radius $R=2GM$.
By plugging (\ref{NS mass-radius}) into this condition,
one obtains the following results:
\begin{eqnarray}
M_{NS} \approx \left ( \frac{1}{2G}\right )^{3/2}m_B^{-1/2}d_0^{3/2}, \qquad 
R_{NS} \approx \left ( \frac{1}{2G}\right )^{1/2}m_B^{-1/2}d_0^{3/2}.
\label{NS mass-radius2}
\end{eqnarray}
Note that we can obtain the moderate values of the neutron star mass $M_{NS}$
and radius $R_{NS}$ 
by using the familiar values of the parameters $m_B, G$ and $d_0$. 
This implies that above very simple analysis
gives good estimation for the  neutron star.
In our case,
combined (\ref{NS mass-radius2}) with the relation $\frac{4}{3}\pi d_0^3 n_{cr} \approx 1$,
one can estimate a  numerical value of 
 $d_0$. Since we have $n_{cr}$ lager than the value of the normal
 nuclear density $0.16\mathrm{fm}^{-3}$ in \cite{GKTTT}, we get the
 values of $M_{NS}$ and $R_{NS}$ which are smaller than previous one. However the suppression from this effect
 is not as strong as it explains our result. It is remained as an open problem.

\par
\vspace{0.5 cm}
(2){\bf the case of $p_c\neq 0$}.
\par 
In this case, the outer core is needed to hold the pressure balance with the inner core of nuclear matter and neutron star has the structure of two phases.
\par
If we assume the outer core is an ideal Fermi gas, $p$ and $\rho$ are expressed by

\bea
  \rho &=&{8\pi\over ( 2\pi )^3}\int_{0}^{k_F} dk~k^2\sqrt{k^2+m^2}\, , \label{energy1} 
\nonumber \\
  p &=&{8\pi\over 3(2\pi )^3}\int_{0}^{k_F} dk~{k^4 \over \sqrt{k^2+m^2}}\, , \label{press1}
 \nonumber
\eea
 with nucleon mass($m=0.94 \rm{ ~GeV}$).

Then, the approximate equation of state is given as
$$p=p_c(\rho/\rho_c)^{\gamma}$$
with $\gamma=5/3$.
\par
\begin{figure}[htbp]
\begin{center}
\includegraphics[width=9.0cm,height=6cm]{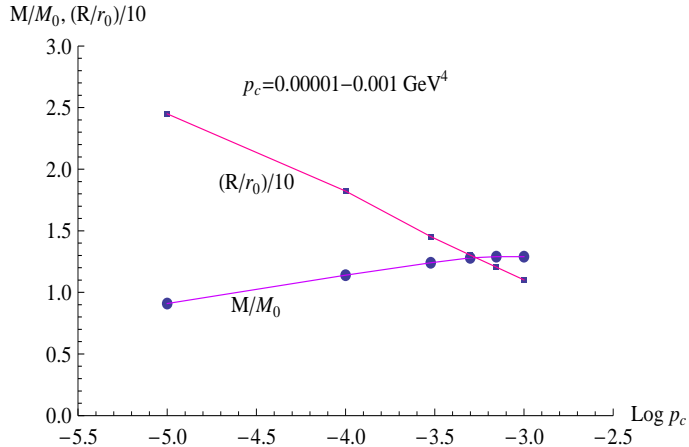}
\caption{{\small The radius and mass of neutron star vs $p_c$
 in 2-phase model. }}
\label{TOV-2}
\end{center}
\end{figure}

We can estimate the radius and mass of neutron star with two phase structure.
The calculated radius and mass are shown in Fig.\ref{TOV-2}  .
\par

When $p_c\geq 10^{-5} \rm{~GeV^4}(\rho_c\geq0.0005 \rm{~GeV^4})$, the radius and mass of the inner core is quite small and the outer core dominates. In case of $p_c\leq 10^{-5}\rm{~GeV^4}(\rho_c\leq0.0005 \rm{~GeV^4})$, inner core of the nuclear matter phase begins to dominate.
Over the wide range of $p_c$, we obtain $R\simeq 10\sim 20 \rm{km}$ and $m\simeq M_s$.

\newpage
\section{Summary and Discussions}

In this paper, based on the previous study of a holographic nuclear matter, we tried to construct
the equation of states (EOS) to obtain the mass-radius relation of neutron stars.
The relation between the chemical potential ($\mu$) and the nucleon number density ($n$) 
leads to the following EOS
$$ p-p_c \propto (\rho-\rho_c)^{\gamma}$$ 
with $\gamma=0.78$, which is less than 1 and quite different from the ordinary ones.
As the nucleon number density increases, the effective nucleon mass and nucleon size also increase in our holographic model. 
This is the origin of  our EOS.
\par
When we apply the nuclear matter to the neutron star,  interesting results about the mass and radius of neutron star are obtained.
In the case of $p_c=0$ and $\rho_c=0$, the neutron star is unstable. 
While in our case, there is a phase boundary with $\rho_c \neq 0$. Due to the property, the neutron star can be stable.
\par
When $p_c\neq 0$, two phase structure is realized. Over the wide range of $p_c$, we obtain the normal neutron star.
\par
In our setup, baryons have been introduced as dilute instanton gas and
the weak attractive force which is needed to form the stable nuclear matter
is realized via the Dirac-Born-Infeld action in the curved background.
 However, the repulsive interaction which has been seen in the two
 baryon system \cite{Hashimoto:2009ys} is absent. 
We have neglected the overlap among instantons in the equation (\ref{dilute-gas-approximation}). 
Because of the absence of the repulsive force, the mass and radius of neutron stars
we obtained here get much smaller than usual.
In the language of  four dimensional field theories, people introduce one pion
exchange as an attractive force between baryons, while omega meson exchange plays a role of
a repulsive core \cite{Schmitt}. 

There are some other aspects of the current work. One of them is the magnetic
field in neutron star. To this end, it is necessary to extend our model
so as to incorporate the electromagnetic fields. This point will be asked in the future.

\section*{Acknowledgements}
K. G thanks A. Nakamura for useful comments.
The work of M. T is supported in part by the JSPS Grant-in-Aid for Scientific Research,
Grant No. 24540280. The work of K.K. is supported by MEXT/JSPS,
Grant-in-Aid for JSPS Fellows No. 25$\cdot$4378. \\



\newpage

\end{document}